\documentclass[twocolumn,aps,preprint,10pt,superscriptaddress,floatfix,showpacs]{revtex4-1}
\usepackage{graphics}
\usepackage{dcolumn}
\usepackage{bm}
\usepackage{epsfig}
\usepackage{color}
\usepackage{epic}
\usepackage{amsmath}
\usepackage{hyperref}

\begin{document}

\title{ Spin reorientation in NdFe$_{0.5}$Mn$_{0.5}$O$_{3}$: Neutron scattering and \emph{Ab-initio} study}



\author{ Ankita Singh}
\affiliation{Department of Physics, Indian Institute of Technology Roorkee, Roorkee 247 667, India
}
\author{A. Jain}
\affiliation{Solid State Physics Division, Bhabha Atomic Research Center, Mumbai, 400 085, India
}
\author{Avijeet Ray}
\affiliation{Department of Physics, Indian Institute of Technology Roorkee, Roorkee 247 667, India
}
\author{Padmanabhan B.}
\affiliation{Department of Physics, Indian Institute of Technology Roorkee, Roorkee 247 667, India
}
\author{S. M. Yusuf}
\affiliation{Solid State Physics Division, Bhabha Atomic Research Center, Mumbai, 400 085, India
}
\author{T. Maitra}
\email{tulimfph@iitr.ac.in}
\author{V. K. Malik}
\email{vivekfph@iitr.ac.in}
\affiliation{Department of Physics, Indian Institute of Technology Roorkee, Roorkee 247 667, India
}


\date{\today}

\begin{abstract}
 The structural, magnetic, and electronic properties of  NdFe$_{0.5}$Mn$_{0.5}$O$_3$ have been studied in detail using bulk magnetization, neutron/x-ray diffraction and first principles density functional theory calculations. The material crystallizes in the orthorhombic $Pbnm$ structure, where both Mn and Fe occupy the same crystallographic site ($4b$). 
 Mn/Fe sublattice of the compound orders in to a G-type antiferromagnetic phase close to 250\,K where the magnetic structure belongs to ${\Gamma}_{1}$ irreducible representation with spins aligned along the crystallographic $b$ direction.  This is unconventional in the sense that most of the orthoferrites and orthochromites order in the ${\Gamma}_{4}$ representation below the N\'{e}el temperature.This magnetic structure then undergoes a complete spin reorientation transition with
temperature in the range 75\,K$\gtrsim$ T $\gtrsim$ 25\,K where the magnetic structure exists as a sum of two irreducible representations (${\Gamma}_{1}$+${\Gamma}_{2}$) as seen from neutron diffraction measurements. At 6\,K, the magnetic structure belongs entirely to ${\Gamma}_{2}$ representation with spins aligned antiferromagnetically along the crystallographic $c$ direction having a small ferromagnetic component ($F_x$). The unusual spin reorientation and correlation between magnetic ground state and electronic structure have been investigated using first principles calculations within GGA+U and GGA+U+SO formalisms.
%
%

\end{abstract}


\maketitle

\section{Introduction}
\label{intro}
Ferromagnetic materials have been utilized extensively for practical applications starting from 200 BC as compass till recently as information storage and read/write devices\cite{stohr2006magnetism,Prinz199957,Prinz199558}. Such versatile applications of ferromagnetic materials have been feasible by virtue of easy manipulation of magnetization utilizing moderate magnetic field\cite{Neel49}. On the other hand, antiferromagnetic ordering is difficult to control through external magnetic field and found limited applications as pinning layer for adjacent ferromagnetic layer in read heads of storage devices\cite{Berkowitz1999552}. Rapid generation of digital data requires enhancement of the data storage density and speed of the data transfer. Later is specifically limited by the time scale of magnetization manipulation ($\sim$10\,ns) due to slow decay and rise time of magnetic fields in present data storage devices.\cite{stohr2006magnetism} Novel methods to manipulate magnetic order using ultra short ($\sim$ps) pules of coherent electromagnetic radiation have unwrapped the possibilities for development of faster storage devices having manipulation time scale of (sub) picosecond scale\cite{stohr2006magnetism,Back1999774}.\\ Rare-earth (RE) orthoferrites, from a family of strongly correlated materials, exhibit two orders of magnitude faster spin dynamics compare to ferromagnetic materials\cite{Kimel:2004p19865,PhysRevB.84.104421}.
 Most of the RE orthoferrites (RFeO$_3$)  are G-type canted antiferromagnets with a weak ferromagnetic component due to Dzyaloshinskii-Moriya (DM) interaction and show temperature induced spin reorientation (SR) from one magnetic symmetry to another magnetic symmetry.  Exchange interactions between Fe$^{3+}$-Fe$^{3+}$,  R$^{3+}$-Fe$^{3+}$ and  R$^{3+}$-R$^{3+}$ play important role to determine the complex magnetic properties (magnetic structure and spin reorientation) of RE orthoferrites. Isotropic Fe$^{3+}$-Fe$^{3+}$ exchange interaction dictates the magnetic anisotropy and hence the  magnetic structure of Fe$^{3+}$ spins below ordering temperature ($T_\mathrm{N}$). Exchange field due to Fe$^{3+}$ moment on R sublattice polarizes the R$^{3+}$ spins. Further, comparatively weaker anisotropic exchange interaction between R$^{3+}$ and Fe$^{3+}$  generate effective fields on  Fe$^{3+}$ spins which in turn start to rotate and try to align perpendicular to R$^{3+}$ spins. This transition might be continuous/abrupt depending upon the of RE element\cite{Yamaguchi}.\\ NdFeO$_3$ is one of the well studied family members of orthoferrite materials which exhibits antiferromagnetic ordering and temperature dependent SR\cite{PhysRev.118.58,epstein1969peakcomp,White69jap,PINTO1972663,sosnowska1982refinement,SOSNOWSK1986394,PRZENIOSLO19952151,przenioslo1995magnetic,PRZENIOSLO1996370,bartolome1997single,slawinski2005spin,Singh08JRS,Chenjap12,yuan2013spin,Chanda20131688,Jiang16jpcm}.
It crystallizes in orthorhombic crystal structure of the space group $Pbnm$  and comprises of highly distorted corner shared Mn(Fe)O$_{6}$ octahedra. NdFeO$_{3}$ is a canted G-type antiferromagnet with a high N\'{e}el temperature ($T_\mathrm{N}$) of 690\,K \cite{slawinski2005spin}. The magnetic structure is represented as G$_x$F$_z$  belonging to irreducible representation ${\Gamma}_{4}$, in which the spins form a G-type antiferromagnetic order along $a$ direction(G$_x$), with a small ferromagnetic component along $c$ direction due to canting (F$_z$). As the temperature is lowered, Fe spins rotate continuously in ac plane in the temperature range 200\,K to 105\,K resulting in a change in the magnetic configuration from G$_x$F$_z$  of ${\Gamma}_{4}$ to F$_x$G$_z$ of ${\Gamma}_{2}$ representation with G-type ordering along the $c$ direction (G$_z$) and a weak ferromagnetic ordering in the $a$ direction (F$_x$) \cite{SOSNOWSK1986394}. Though the whole process of SR is quite complex, a noticeable ordering of Nd moments as C-type antiferromagnet emerges at 1.5\,K\cite{bartolome1997single, przenioslo1995magnetic,PRZENIOSLO1996370}. The weak ferromagnetic moments, F$_x$ of Fe and f$_x$ of Nd, in NdFeO$_3$ single crystals are antiparallel and  the overall spontaneous magnetization is compensated at T = 8\,K\cite{yuan2013spin}. NdFeO$_3$ could be utilized in high frequency electronic components  owing to its large band gap of about 2.5\,eV.\\
 Another RE transition metal oxide which has drawn considerable attention for its intriguing properties is NdMnO$_{3}$. NdMnO$_3$ is A-type anti-ferromagnet with relatively low $T_\mathrm{N}$ of 75\,K. Although NdMnO$_3$ and NdFeO$_3$ are isostructural and antifrromagnetic, but underlying mechanism responsible for their complex magnetic properties is quite different. In NdMnO$_{3}$, the Mn ion has 3$+$ valence state with $3d^{4}$ electronic configuration, which  splits under crystal field into low lying triply degenerate t$_{2g}$ states and doubly degenerate e$_{g}$ states at higher energies \cite{rasera1998mn, jandl2005infrared}. 
The doubly degenerate e$_{g}$ is split into e$_{g}^{1}$ and e$_{g}^{2}$ states due to the Jahn-Teller distortion which is also manifested as structural distortion. This results in tilting of octahedra by ${\sim}$ 45$^{o}$ in a zig-zag manner and also compression of the in-plane Mn-O bonds and elongation of the apical Mn-O bond.
  As a consequence of crystal field splitting and co-operative Jahn-Teller (J-T) effect, the highest occupied e$_{g}$ orbital orders as {$d_{{3x}^{2}-{r}^{2}}$ and $d_{{3y}^{2}-{r}^{2}}$ alternately at adjacent Mn sites in the $a$-$b$ plane \cite{chatterji2009direct}. Such a staggerred in-plane orbital ordering leads to a ferromagnetic alignment of spins in $a$-$b$ plane and antiferromagnetic alignment along the $c$ direction resulting in A-type AFM with  $T_{N}$ at around 75\,K \cite{wu2000magnetic}. As temperature is further reduced, 
  Nd sublattice orders ferromagnetically along the $a$ direction  below 20\,K the \cite{chatterji2009direct}. The electronic properties of this system falls in an intermediate regime between a Mott insulator and charge transfer insulator with a large band gap of about 1\,eV driven primarily due to J-T distortion. This has been studied from density functional theory based band structure calculations.\cite{Balasubramanian201450}\\
In view of the very different magnetic properties shown by NdFeO$_3$ and NdMnO$_3$, it would be extremely interesting to look at an intermediate system where we  substitute half of the Fe ions with Mn. In this situation, one would expect a very complex interplay of spin, lattice, orbital degrees of freedom between the Mn and Fe sublattices and hence a new set of enriching physical properties.
Substitution of Mn$^{3+}$ ions at the Fe site not only makes it J-T active, it also tunes the super exchange interaction, influences the magnetic symmetry and spin reorientation through the Mn-Fe, Nd-Mn and Nd-Fe interactions\cite{zhang2010first,mihalik2013magnetic,yadav2015magnetic}. The electron-phonon coupling will also change due to change in J-T interaction upon doping. \\
In the present work, we have studied the effect of 50\,\% Mn substitution at Fe sites on the magnetic structure, spin reorientation, and electronic structure of NdFe$_{0.5}$Mn$_{0.5}$O$_3$ (NFMO). We have used various experimental techniques such as x-ray diffraction, magnetization, neutron diffraction, neutron depolarization, and also theoretical techniques such as density functional theory calculations to establish the complex magnetic structure of NFMO compound as well as to understand the underlying mechanism. A unique magnetic symmetry $\Gamma_1$ was observed below ($T_\mathrm{N}$) which transforms  gradually into a commonly observed symmetry $\Gamma_2$ through a SR transition with temperature.  To the best of our knowledge, the $\Gamma_1$ symmetry has not been observed previously above spin reorientation transition temperature in the family of orthoferrites. The Observation of two magnetic structures and spin reorientation in NFMO has been explained on the basis of magnetic anisotropy energy calculated using first principles calculations. 

\section{Methods}
\label{methd}
\subsection{Experimental}
\label{Expmethd}
Powder sample of NFMO was  synthesised using standard solid state reaction method. Nd$_{2}$O$_{3}$, MnO$_{2}$ and Fe$_2$O$_3$ were weighed according to appropriate stoichiometry and grounded in an agate mortar for 12 hours. The heat treatment of the sample was carried out according to procedure explained in \cite{C5CP07083J}. Structural phase of the sample was  identified using a  Bruker D8 two circle x-ray diffractometer at Cu $K_{\alpha}$ wavelength. The stoichiometry of Nd and transition metals was confirmed using energy dispersive x-ray analysis (EDAX) on the sample pellet. Average magnetization measurements were performed using vibrating sample magnetometer (VSM) Quantum Design physical property measurement system (PPMS). Further measurements were repeated using superconducting quantum interference device (SQUID) of Quantum Design magnetic measurement system (MPMS) to confirm the results obtained from VSM of PPMS. Zero field cooled (ZFC) and field cooled (FC) measurements from 350\,K to 10\,K in 0.01\,T field were carried out to identify the different magnetic transitions and their respective temperatures. In addition magnetization vs temperature was also measured between 300\,K and 500\,K using an additional vibrating sample magnetometer in a field of 0.2 T to find out the effective paramagnetic magnetization. Field variation of magnetization was carried out at various temperatures between 300 and 5\,K. Neutron diffraction studies in absence of magnetic field were carried out at various temperatures between 4 to 300\,K for obtaining the temperature variation of crystal structure as well as magnetic structure. Neutron diffraction measurements were carried out at powder diffractometer (PD)-I ($\lambda = 1.094\,\mathrm{\AA}$) and II ($\lambda = 1.2443\,\mathrm{\AA}$), the Dhruva reactor, Trombay, Mumbai, India.  The diffraction data were analyzed using FullProf\cite{rodriguez1990fullprof} suite of programs24 employing the
Rietveld method \cite{rietveld1969profile}. Magnetic structure was determined using the irreducible representations from BASIREPS\cite{rodriguez2011basireps} and refined using FullProf. One dimensional neutron depolarization measurements over the temperature range of 4-300\, K, to identify the short range weak ferromagnetism,  were carried out using the polarized neutron spectrometer (PNS) also at the Dhruva reactor, under an applied magnetic field of 50\,Oe. For the neutron depolarization measurements, polarized neutrons ($\lambda = 1.205\,\mathrm{\AA}$) were produced and analyzed by using the (111) reflection of magnetized Cu$_2$MnAl Heusler single crystals. The two different states (up and down) of the incident neutron beam polarization was achieved by a $\pi$-fipper just before the sample. The polarization of the neutron beam was determined by measuring the intensities of neutrons in non-spin flip and spin flip channels with the flipper off and on (flipping ratio), respectively.
\subsection{Theoretical}
\label{Theomethd}
Electronic structure of NFMO was obtained using the projector-augmented wave(PAW) psuedopotential and a plane wave basis method within the density functional theory framework as implemented in the Vienna ab-initio simulation program (VASP)\cite{kresse1996efficient}. Calculations are performed within Perdew-Burke-Ernzerhof generalized gradient approximation (PBE-GGA)\cite{perdew1996generalized} and GGA+U\cite{anisimov}. Non-collinear magnetic calculations were performed within GGA+U+SO approximation. The Mn(Fe) $3d, 4s$, O $2s, 2p$ and Nd$5p, 5d, 6s$ were treated as valence states. An energy cut-off of 400 eV was used for the plane wave basis set while a 6x6x6 Monkhorst-Pack $k$-mesh centered at ${\Gamma}$ was used for performing the Brillouin zone integrations. Ionic positions were relaxed until the forces on the ions are less than 0.1 meV/$\mathrm{\AA}$. The calculations were performed for different magnetic configurations viz. ferromagnetic, A, C and G type antiferromagnetic configurations.
\begin{figure}[h!] \center
       \begin{picture}(240,150)
        \put(-5,-5){\includegraphics[width=230\unitlength,]{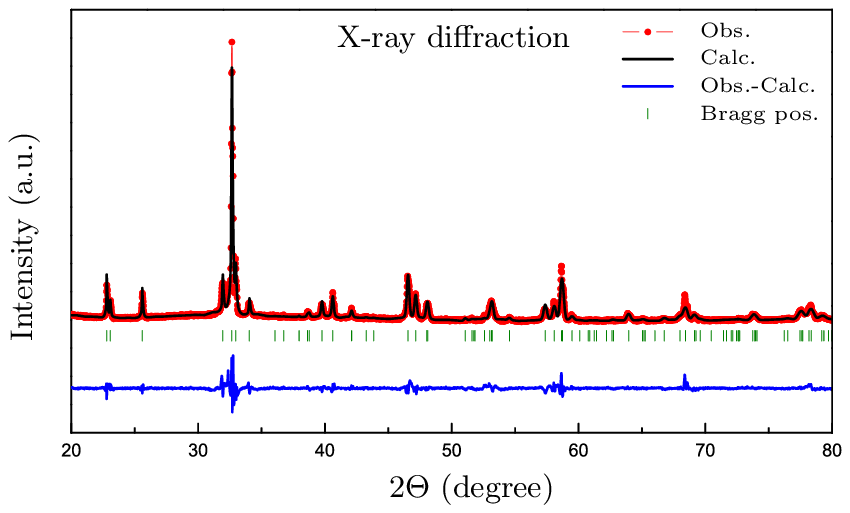}}
          
                    
       \end{picture}
\caption{Experimental and refined x-ray diffraction pattern of NFMO at room temperature. }
\label{xrd}
\end{figure}
\begin{figure}[h!] \center
       \begin{picture}(240,135)
        \put(-5,-5){\includegraphics[width=230\unitlength,]{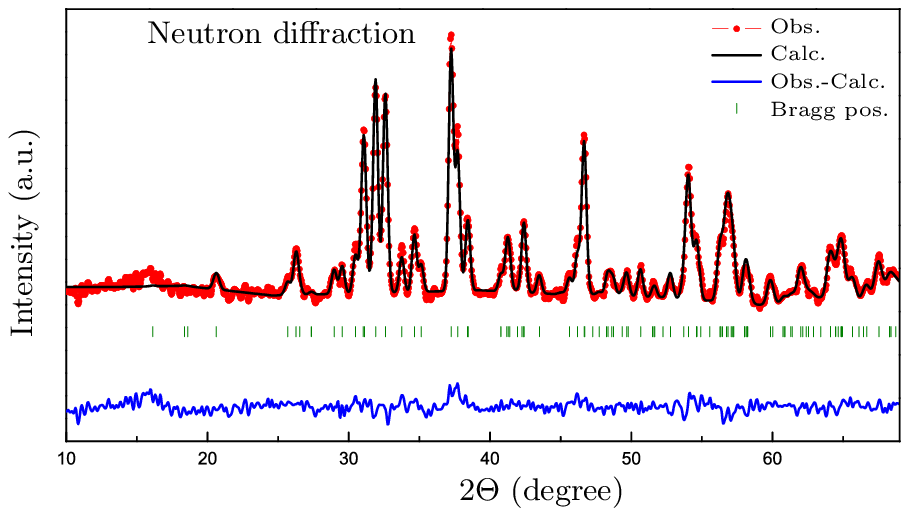}}
          
                    
       \end{picture}
\caption{Experimental and refined x-ray diffraction pattern of NFMO at room temperature. }
\label{nd_300}
\end{figure}
\section{Experimental results}
\label{res}
\subsection{Structural Characterization}
\label{stuct_char}
Fig.~\ref{xrd} shows the room temperature x-ray diffraction pattern of the sintered NFMO powder sample. Orthorhombic structure of the sample has been considered for refinement shown in Fig.~\ref{xrd}. 
The preliminary analysis of the x-ray diffraction data confirms phase formation and purity of the NFMO sample. The mixed doped NFMO compound can be considered as double perovskite Nd$_{2}$MnFeO$_{6}$ with B-site ordering of Mn and Fe ions. B-site ordering of the compound should mark the unique positions for transition metal ions which results in monoclinic structure. Random occupation of B site by transition metal ions leads to orthorhombic crystal structure of NdFeO$_3$ and NdMnO$_3$.  The diffraction pattern was refined to orthorhombic ($Pbnm$) as well as monoclinic ($P2_{1}/n$) spacegroups with comparable $\chi^2$ values. Hence, it is not possible to confirm the crystal structure using x-ray diffraction. However our attempts to find traces of weak  (021) Bragg peak, which should occur only in monoclinic structure, did not succeed. Besides the observed magnetic behavior is entirely different from that expected from a double perovskite compound which predominantly show ferromagnetism e.g. in R$_{2}$MnBO$_{6}$ (R=La,Nd and B=Ni,Co) \cite{yadav2015magnetic, anshul2014magnetodielectric,PhysRevB.68.064415,PhysRevB.67.014401}. This information indicates the possible phase formation of NFMO with orthorhombic symmetry. However the structure could be confirmed using neutron diffraction measurements. Unlike x-ray diffraction, form factors of Mn and Fe are not similar in case of neutron diffraction. This distinction helps to identify the B site order/disorder. Fig.~\ref{nd_300} shows measured and refined neutron diffraction pattern of NFMO for 300\,K at PD-II using neutron with wavelength of  1.205\,$\mathrm{\AA}$. The diffraction pattern was fitted to orthorhombic structure as shown in Fig.~\ref{nd_300}, which shows excellent agreement with the experimental pattern. However the monoclinic structure gives a much poorer quality of fitting (not shown) unlike the x-ray pattern. Thus it can be concluded that NFMO crystallizes in the orthorhombic space group and the Mn/Fe atoms are arranged in a random manner at the $4b$ Wyckoff positions.\\
The neutron diffraction patterns were collected systematically at regular intervals between 4\,K to 300\,K. In this section, the structural analysis of NFMO in comparison to NdFeO$_3$ and NdMnO$_3$ is dicussed. 
%
\begin{table*}
\caption{Structural parameters of NdMn$_{0.5}$Fe$_{0.5}$O$_{3}$. }
\begin{center}
\begin{tabular}{p{4.0cm}c c c c c c c }\hline\hline 
 & \multicolumn{3}{c}{} & \multicolumn{1}{c}{} & \multicolumn{1}{c}{}\\
Compound & \multicolumn{3}{c}{NdMn$_{0.5}$Fe$_{0.5}$O$_{3}$ } &  \multicolumn{1}{c}{ NdMnO$_{3}$}&\multicolumn{1}{c}{NdFeO$_{3}$}\\ \hline 
Parameters & 300 K  & 100 K & 6 K  & 295 K  & 295 K \\ &&&&&\\  \hline
a({\AA}) & 5.4235 & 5.4149  & 5.4145 & 5.4170     & 5.4510 \\
b({\AA}) &  5.5977 & 5.6003 & 5.5999 &5.8317   & 5.5880 \\  
c({\AA}) & 7.6929 & 7.6710 & 7.6697&7.5546  & 7.7616 \\ 
Mn(Fe)-O(1)(m){\AA} & 1.980 & 1.979 & 1.978 &1.951    & 2.0012\\  
Mn(Fe)-O(2)(l){\AA} & 2.050 & 2.053 & 2.054 & 2.218   & 2.0226\\
Mn(Fe)-O(2)(s){\AA} & 1.975 & 1.968 & 1.965 &1.905  & 2.0072\\ 
Mn(Fe)-Mn(Fe)(p){\AA} & 2.050 & 2.053 & 2.054 & 3.9342   & 3.8810\\
Mn(Fe)-Mn(Fe)(o){\AA} & 1.975 & 1.968 & 1.965 & 3.7945 & 3.9031\\
\hline\hline
\end{tabular}
\end{center}
\label{Table_1}
\end{table*}
In Table~\ref{Table_1}, the lattice constants along with M-O(M = Fe, Mn) bond lengths and M-O-M bond-angles of NdFe$_{0.5}$Mn$_{0.5}$O$_{3}$ have been shown for 300K, 100K and 6\,K. For comparison, we have included the corresponding details of NdMnO$_{3}$ and NdFeO$_{3}$ from earlier studies\cite{munoz2000magnetic, slawinski2005spin}. We observe that the lattice parameters ($a$ and $c$) of NdFeO$_{3}$ are greater than those of NFMO while $b$ of NdMnO$_{3}$ is greater than that of NFMO. Thus the lattice parameters of NFMO are intermediate, though we find that the values are closer to that of NdFeO$_{3}$ ($\Delta a=-0.51\%, \Delta b=+0.17\% ,\Delta c=-0.89\%$) rather than the parent manganite NdMnO$_{3}$  ($\Delta a=+0.12\%, \Delta b=-4\% ,\Delta c=+1.83\%$). We note the large changes in $b$ and $c$ lattice parameters in comparison to  NdMnO$_{3}$. These values are also in good agreement with recent work done by T. Chakraborty et al. on a series of NdMn$_{1-x}$Fe$_x$O$_3$ samples\cite{C5CP07083J}. In case of RMn$_{1-x}$Fe$_x$O$_3$(R=RE element) it has been observed that the long range cooperative (static) J-T distortion is effective for lower values of x ($\approx 0.2\sim0.3$) and structure is classified as O$'$-orthorhombic with $a<c/\sqrt{2}<b$ in this regime. Static J-T distortion disappears at higher values of x and structure changes into O-orthorhombic  with $c/\sqrt{2}<a<b$. A weaker orbital ordering associated with dynamical J-T effect at Mn sites may still be present in the system up to x$\approx0.5$\cite{Hong11APL,hong2012dielectric,C5CP07083J}.  This inequality $a <c/{\sqrt 2}<b$  is also observed in case of NFMO with a small difference between $a$ and $c/{\sqrt 2}$. 
\\
Three different M-O bond lengths have been listed in Table~\ref{Table_1}.  Long ($l$) and short ($s$) bond lengths correspond to the M-O(2) bonds in the $a-b$ plane. Medium ($m$) bond length corresponds to the out of plane M-O(1) apical bond which is almost parallel to the $c$ axis. In NdFeO$_{3}$, $l$, $m$ and $s$ are almost equal implying that the distortion of the octahedra is minimal. While in NdMnO$_{3}$, the bond lengths are highly unequal due to the J-T distortion associated with orbital ordering in the $a-b$ plane.
In NFMO, it is expected that the Mn-O and Fe-O bond lengths should be unequal. However,  the only average bond lengths for the three M-O bonds are obtained experimentally. Comparing the M-O bond lengths of NFMO with NdFeO$_{3}$, a slight decrease in $s$ and an increment in  $l$, $m$ is observed. Though the changes are minimal it indicates possibility of the J-T distortion in the $a-b$ plane.
The in-plane and out of plane distortions are characterized by the Jahn-Teller parameters $Q_{2}$ = 2($l-s$)/${\sqrt 2}$ and $Q_{3}$=2($2m-l-s$)/${\sqrt 6}$ respectively\cite{Alonso2000}.
Values of  $Q_{2}$ and $Q_{3}$ are nearly zero in case of NdFeO$_3$ and 0.1064 and -0.0526 for NFMO as obtained using bond lengths from Table~\ref{Table_1}. In case of NFMO the J-T parameters are clearly non zero, but are much smaller compared to NdMnO$_{3}$ ($Q_{2}$ =  0.4426, $Q_{3}$ =  -0.1797).

An additional parameter often used to quantify the distortion of the octahedra is ${\Delta}_{d}$ defined as ${\Delta}_{d}$=(1/6)${\Sigma}_{n=1,6}[(d_{n}-<d>)/<d>]^{2}$\cite{Alonso2000} where d$_{n}$s are the individual M-O bond lengths.  
 In case of NFMO, we observe that the level of octahedral distortion (10$^{4}$${\Delta}_{d}$=3.5) is smaller by an order of magnitude in comparison to NdMnO$_3$ (10$^{4}$${\Delta}_{d}$=46.4)  and quite close to NdFeO$_3$ (10$^{4}$${\Delta}_{d}$=0.2). Due to the presence of such high level of octahedral distortions long-range static J-T effect and orbital ordering are seen to exist in NdMnO$_3$ \cite{C5CP07083J}.\\
Also, the temperature variation (not shown) of $a$ and $c$ lattice parameters is quite similar to that of NdFeO$_{3}$. However the parameter $b$ shows a continuous increase with decreasing temperature, unlike in NdFeO$_{3}$ which shows an initial decrease till 150\,K and then continuously rises till 4\,K \cite{slawinski2005spin}. The increase in $b$ can also be linked with the variation of the M-O(2)(l) bond lengths with temperature and possible increase in J-T ditortion at Mn sites.

Presence of a dynamical J-T effect associated with a weak and short range orbital ordering in NFMO could be inferred from the above mentioned facts about lattice constants, bond lengths, J-T parameter and temperature dependent data. 
  \begin{figure}[t!] \center
       \begin{picture}(250,160)
        \put(0,-5){\includegraphics[width=240\unitlength]{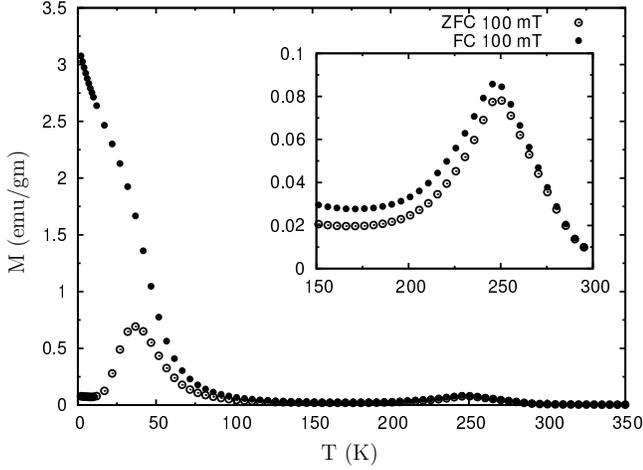}}
          
                    
       \end{picture}
        \caption{ ZFC-FC plots of NFMO from 2 to 350 K showing N\'{e}el temperature at $T_\mathrm{N}$ = 250\,K and weak ferromagnetism below 70\,K.}
        \label{ZFC_FC}
      \end{figure}
\subsection{Magnetic properties}
\label{Magnetic_prop}
\subsubsection{Average magnetization}
\label{avg_mag}
As shown in Fig.~\ref{ZFC_FC}, zero field cooled (ZFC) and field cooled (FC) magnetization measurements were performed from 2\,K to 350\,K in the magnetic field of 100\,mT. Temperature dependent ZFC and FC show a characteristic cusp (see inset of Fig.~\ref{ZFC_FC}) at 250\,K denoting antiferromagnetic ordering of the M (Mn/Fe) spins. Parent compound NdFeO$_3$ undergoes from paramagnetic state to G-type antiferromagnetic state at 690\,K\cite{slawinski2005spin}. As recent reported by Chakraborty et al. \cite{C5CP07083J}, $T_\mathrm{N}$ is seen to decrease systematically with increase of Mn substitution in NdFe$_{1-x}$Mn$_x$O$_3$.  Value of  $T_\mathrm{N}$ is 622\,K for x=0.1, 529\,K for x=0.2, 449\,K for x=0.3 and 356\,K for x=0.4 in NdFe$_{1-x}$Mn$_x$O$_3$\cite{1742-6596-592-1-012117}; therefore, observed value of $T_\mathrm{N}$ (250\,K) for NFMO is consistent with previous reports. From the Curie-Weiss temperature ${\theta}$ = -401 K and $T_{N1}$ = 250 K, the nearest neighbor and next nearest neighbor interactions are estimated to be $J_{1}$ = -0.95 meV and $J_{2}$  = -0.11 meV respectively in the mean field approximation\cite{Tsushima}.
 Susceptibility does not follow Curie-Weiss law above 250\,K in the whole high temperature range as expected from a paramagnetic phase. Such behavior suggests existence of short range correlations above $T_\mathrm{N}$ which could also be inferred from a broad hump observed approximately at 16$^\circ$ in room temperature neutron diffraction data  shown in Fig.~\ref{nd_300}.\\ 
 ZFC and FC magnetization exhibit an increase below 70\,K and a large bifurcation below 50\,K. ZFC magnetization decreases below 50\,K whereas FC magnetization keeps increasing with a slight change in slope at 50\,K. Such magnetization is observed typically for spin glass systems, weak ferromagnetic systems and magnetism with short range ordering.\cite{blundell2001magnetism} FC magnetization does not follow the mean field magnetization for a ferromagnet with long range ordering. Troyanchuk et al.\cite{Troyanchuka} attributed the low temperature magnetic behavior in NFMO entirely to spin glass state. However, frequency dependent ac-susceptibility measurements ruled out presence of spin glass state in the present study. Further magnetization (M) vs magnetic field (H) loops were measured at different temperature to understand the low temperature magnetization (below 70\,K) of NFMO.\\
\begin{figure}[t!] \center
     \begin{picture}(240,330)
        \put(-5,-90){\includegraphics[width=250\unitlength,]{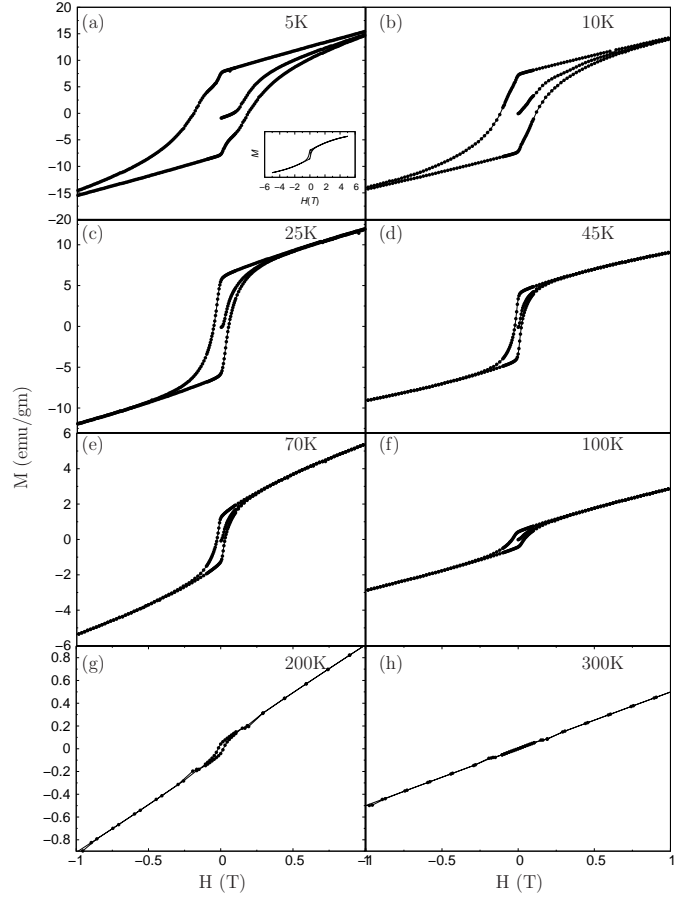}}
          
                    
       \end{picture}
\caption{M-H plots of NFMO at (a) 5K (b) 10K (c) 25K (d) 45K (e) 70K (f) 100K (g) 200K (h) 300K }
\label{nfmo-mh}
\end{figure}
  \begin{figure}[b!] \center
       \begin{picture}(250,160)
        \put(0,-15){\includegraphics[width=240\unitlength]{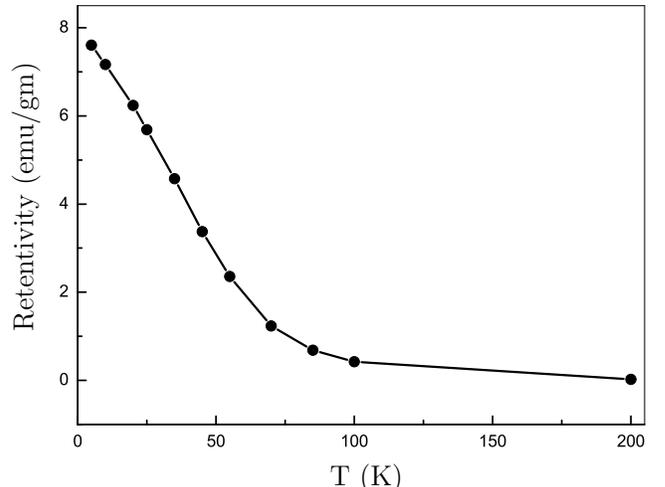}}
          
                    
       \end{picture}
        \caption{Variation of retentivity with temperature in NFMO}
        \label{ret_NFMO}
      \end{figure}
 The magnetization isotherms (M-H loops) of NFMO at various temperatures from 5\,K till 300\,K have been shown in Fig~\ref{nfmo-mh}. At 300\,K, the magnetization shows a linear behavior without any hysteresis indicating as expected from a majority paramagnetic phase. At 200\,K and 100\,K, the M-H loops are quite similar having a deviation from the linear behavior with a small coercivity, retentivity and unsaturated magnetization which is usually seen in antiferromagnetic materials. Nature of M-H loops below 70\,K changes considerably (both qualitatively and quantitatively) with an increase in coercivity and retentivity. Another qualitative change in M-H loop occurs below 30-25\,K with a large coercivity and retentivity value. Shape of the M-H loops obtained at 10 and 5\,K  is significantly different from loops measured at 25, 45 and 70\,K. \\
 
The M-H loops below 100\,K indicate development of a ferromagnetic component. However, the magnetization does not undergo saturation even at 5\,K in a field of 5\,T as seen in the inset of Fig.~\ref{nfmo-mh}(a) indicative of a ferromagnetic component superimposed over antiferromagnetic phase. We present the variation of retentivity versus temperature in Fig.~\ref{ret_NFMO}. The Remanent magnetization is seen to increase below 100\,K suggesting a development of some kind of magnetic ordering. 
As the spin-reorientation effects are characteristic features of the orthoferrite compounds in which the spins rotate from one crystallographic direction to another in a continuous manner, indicative of a second order phase transition \cite{nair2016magnetic, nhalil2015spin}, such a scenario can also provide a 
possibile explaination to the magnetization data of NFMO presented above. However, it is very difficult to conclude about the type of magnetic structure and nature of spin reorientation based on the average magnetization data. Transition occurs most probably from one type of antiferromagnetic order to another type having a weak ferromagnetism below 100\,K. Further microscopic analysis of magnetic structure is required to completely understand the magnetic properties NFMO which we discuss in the following section. 
\\

\subsubsection{Microscopic magnetization}
\label{neutron_diff}
  \begin{figure}[h!] \center
       \begin{picture}(250,140)
        \put(0,-5){\includegraphics[width=240\unitlength]{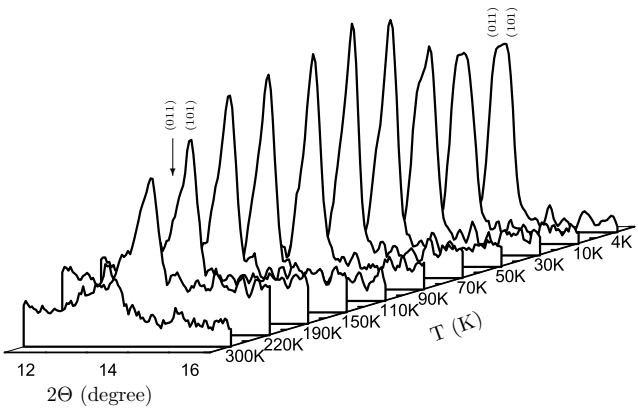}}
          
                    
       \end{picture}
        \caption{Evolution of magnetic peak in neutron diffraction data as a function of temperature}
        \label{ND_temp}
      \end{figure}
In this section, a systematic evolution of the NFMO magnetic structure based on  neutron diffraction data has been discussed. In Fig.~\ref{nd_300} we have already seen the  neutron diffraction pattern at 300\,K, with a broad hump corresponding to short range magnetic ordering near 2${\Theta}$ = 16$^{\circ}$($\lambda = 1.2443\,\mathrm{\AA}$), which confirms that the magnetic correlations exist even above $T_\mathrm{N}$.\\
We now present in Fig.~\ref{ND_temp}, the temperature dependent neutron diffraction measurements which have been performed at PD-II ($\lambda =1.094\,\mathrm{\AA}$).  As temperature is lowered below 250\,K,  a clear magnetic Bragg peak is observed at ${\sim}$ 14.2$^{\circ}$ which is convolution of (011) and (101) peaks. The nuclear Bragg reflections due to the two peaks are forbidden in the $Pbnm$ space group. However, their magnetic reflections have non-zero structure factors. It is not possible to completely resolve both the peaks at incident neutron wavelength of 1.094\,${\mathrm\AA}$ due to resolution limit of the instrument. This further confirms 250\,K as the value of $T_\mathrm{N}$, which is also suggested by magnetization measurements in earlier section.  Intensity of the magnetic peak increases and (011) peaks occurs at a lower angle in the form of shoulder hump on the total magnetic Bragg peak  down to 70\,K. Ratio of the two peaks ($I(011)/I(101)$) is approximately 3 at 70\,K suggesting a G-type antiferromagnetic structure with spins along the $y$ direction\cite{epstein1969peakcomp}. Spectral weight of (101) peak shifts to (011) peak below 70\,K and ratio of both peaks become equal close to 4\,K which represents a G-type antiferromagnetic structure with spins along $z$ direction\cite{epstein1969peakcomp}. Such temperature induced changes in the magnetic peaks clearly stipulate a spin reorientation below 70\,K. Further detailed analysis of the magnetic peaks has been performed to quantify the magnetic structure and temperature induced spin reorientation in NFMO. 
  \begin{figure}[b!] \center
       \begin{picture}(260,500)
        \put(-15,-10){\includegraphics[width=260\unitlength,]{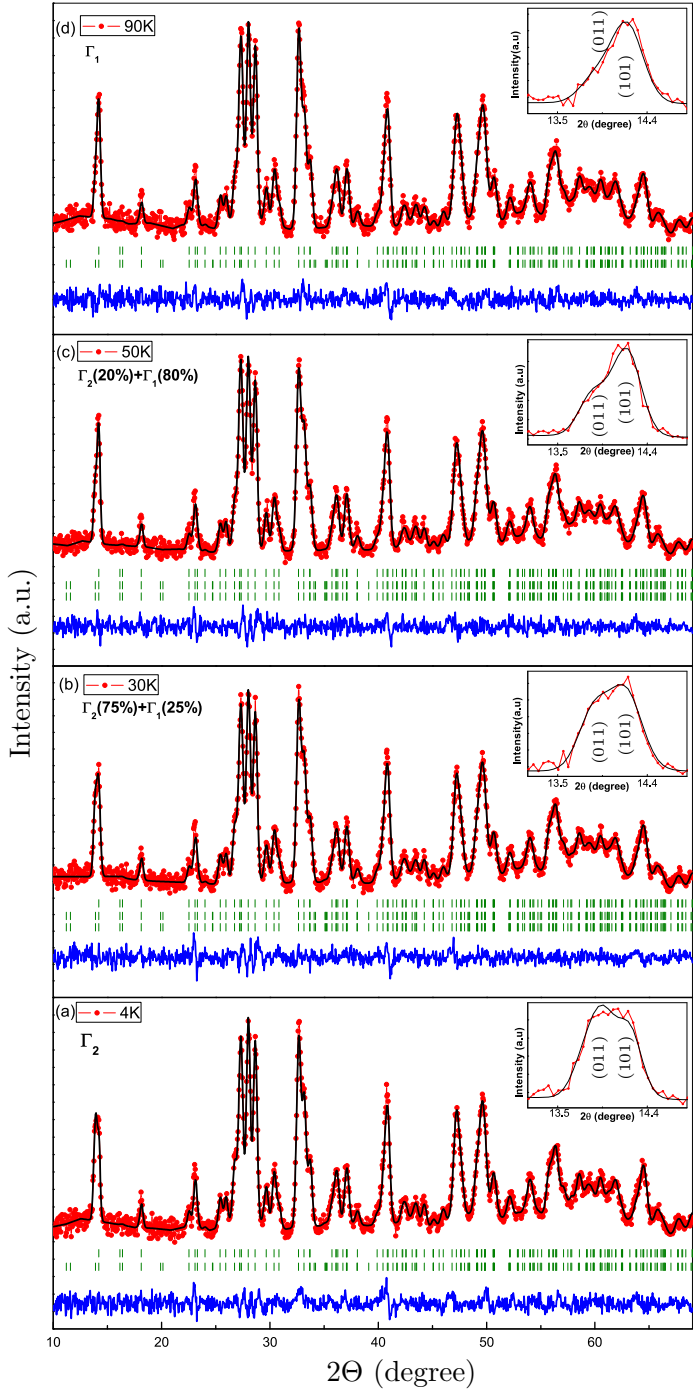}}
          
                    
       \end{picture}
    \caption{Neutron diffraction pattern of NFMO for various temperatures from 4\,K to 300\,K.}
        \label{n_diff}
      \end{figure}
The ordering vector in NFMO is k = (0,0,0), which is same for the two end compounds, indicating that the magnetic and structural unit cells are identical. Comparison of the diffraction pattern with the two end compounds suggest that the antiferromagnetic ordering is identical with G-type NdFeO$_{3}$ rather than the A-type NdMnO$_{3}$. 
To obtain the detailed spin configuration in a unit cell, refinement of the magnetic structure  has been performed for the data collected at 90\,K, 50\,K, 30\,K, and 4\,K. The four temperatures cover completely the reorientation transition region. 
 Since Mn and Fe occupy the same $4b$ site in NFMO, the Fe-Mn ordering is identical. On the other hand, Nd occupies $4c$ and thus can order independently, however in conjunction with the Fe-Mn ordering.\\
For the experimentally determined magnetic ordering vector k=(0,0,0), there exists eight irreducible representations, ${\Gamma}_{1}$ to ${\Gamma}_{8}$.  Four out of these eight representations correspond to zero coefficients. 
Thus we consider four irreducible representations ${\Gamma}_{1}$ to ${\Gamma}_{4}$ which correspond to the Shubnikov magnetic space groups, ${\Gamma}_{1}$ ($Pbnm$), ${\Gamma}_{2}$ ($Pbn'm'$), ${\Gamma}_{3}$ ($Pb'nm'$) and ${\Gamma}_{4}$ ($Pb'n'm$). Using Bertraut's notation\cite{bertaut1963magnetism}, the four magnetic space-groups can be written in a simplified manner as $A_xG_yC_z$, $F_xA_yG_z$, $C_xF_yA_z$ and $G_xC_yF_z$ respectively corresponding to magnetic ordering of the cartesian components of the four $M^{3+}$ spins in the unit cell. A and C type antiferomagnetic ordering is due to hidden canting of the spins, while $F_{x/y}$ is due to overt canting of spins corresponding to ferromagnetic component. Both antiferromagnetic types of the ordering are not observed experimentally in present study probably due to small magnitude of the overt and hidden angles and  have not been discussed further. \\
Corresponding to ${\Gamma}_{1}$,${\Gamma}_{2}$, ${\Gamma}_{3}$ and ${\Gamma}_{4}$ representations, Nd spins can order in $C_z$, $F_xC_y$, $C_xF_y$ and $F_z$ respectively\cite{nair2016magnetic}.
However in all our refinements, inclusion of the symmetry representations of Nd ordering result in poorer quality of fitting due to which we only consider the $M$ sublattice ordering with temperature.\\
At 90\,K (Fig.~\ref{n_diff}a), the best fit is obtained for ${\Gamma}_{1}$ representation, \emph{i.e.} the $G_y$ structure. The spins point in the $y$ direction and the four $M$ atoms in the unit cell arrange as $G$-type antiferromagnet. Also we do not observe any ferromagnetic  component due to canting of the spins in accordance with ${\Gamma}_{1}$ representation. It should be noted that this is first report of antiferromagnetic ordering with ${\Gamma}_{1}$ representation above spin reorientation transition temperature (high temperature phase) in orthoferrites. In almost all the orthoferrites and orthochromites, the first magnetically ordered phase belongs to ${\Gamma}_{2}$ or ${\Gamma}_{4}$ representations.
 However in NFMO, below $T_\mathrm{N}$, the magnetic structure belongs to the ${\Gamma}_{1}$ representation with the y-direction as the easy axis,  unlike the usual trend observed in orthoferrites (pure as well as substituted ones (at R or M sites)). Few of the orthoferrites, for instance Dy$_{0.5}$Pr$_{0.5}$FeO$_{3}$ attain this structure in the low temperature after an abrupt spin-reorientation transition from ${\Gamma}_{2}$ or ${\Gamma}_{4}$.\cite{Yamaguchi, hailongwu14}. In TbFe$_{0.5}$Mn$_{0.5}$O$_{3}$, the ${\Gamma}_{1}$ phase develops as a co-existing phase with the high temperature ${\Gamma}_{4}$ phase. Similarly in DyFeO$_{3}$, the ${\Gamma}_{1}$ phase is attained below 35\,K. However in DyFe$_{0.5}$M$_{0.5}$O$_{3}$, there occurs an abrupt reorientation transition from ${\Gamma}_{4}$ to ${\Gamma}_{1}$ phase at 250\,K. 
 Thus the ${\Gamma}_{1}$ phase can be attributed to the large single ion anisotropy of Dy and Tb based orthoferrite compounds which causes the abrupt reorientation. However in NFMO, the ${\Gamma}_{1}$ phase seems to be the preferred magnetic ordering at high temperature (above spin reorientation temperature) itself. Since the single-ion anisotropy of Nd is much smaller and could be neglected in this high temperature region, the development of ${\Gamma}_{1}$ phase can be attributed to the single ion anisotropy of Mn ion. The   Hamiltonian corresponding to single ion anisotropy term of Mn/Fe ions is given by $DS_{z}^{2}$-$E(S_x^2-S_y^2)$, where D and E are anisotropic constants. The necessary condition for stability of ${\Gamma}_{1}$ structure is $E>$ 0 and -$D>E$ \cite{Yamaguchi}.\\
 Though Fe and Mn ions are in the high-spin 3+ state the total wavefunction of Fe$^{3+}$ ($S$=5/2) has $A_{1g}$ symmetry which is highly isotropic. On the other hand, in Mn$^{3+}$ ($S$=2), the total ground state wavefunction is $E_{g}$-type in which the $d$ orbitals have anisotropic shape. From electron spin resonance studies of transition metal octahedral complexes, it is found that the values of D and E in Mn$^{3+}$-complexes are almost two orders of magnitude greater than that in Fe$^{3+}$-complexes.\cite{Gerber}Thus the easy axis of magnetization of Fe/Mn sublattice is decided by the large single ion anisotropy of Mn$^{3+}$ ion.\\
%
At 50\,K (Fig.~\ref{n_diff}a), an increase in the ratio of intensities of the (011)/(101) peaks is observed due to the ongoing re-orientation. 
At this temperature, the magnetic structure is best refined by a mixture of ${\Gamma}_{1}$and ${\Gamma}_{2}$, wherein the two phases exist in a ratio of 80:20 respectively. The ${\Gamma}_{2}$ representation, corresponding to $F_xG_z$ spin structure  has the $z$ component as the major spin component. Similar to ${\Gamma}_{1}$, the $M$ sublattice spins also order as G-type in $\Gamma_{2}$ representation. However, a weak ferromagnetic component in x direction ($F_x$) is also obtained in the refinement which is associated with $\Gamma_{2}$ representation. Such ferromagnetic component also explains increment in the width of hysteresis loop from magnetization isotherms.  The diffraction pattern is best fitted to combination of ${\Gamma}_{1}$ and ${\Gamma}_{2}$ occurring in the ratio  of 25:75 (Fig.~\ref{n_diff}c) at 30\,K.\\ 
Finally as shown in Fig.~\ref{n_diff}d, the (011) and (101) magnetic peaks show a prominent separation with I(011)/I(101) ${\sim}$ 1 at 4\,K. This also also confirms the fact that its the $z$-component of spins that are aligned as G-type antiferromagnet. Refinement as shown in  Fig.~\ref{n_diff}d confirms that the magnetic structure entirely belongs to ${\Gamma}_{2}$ representation at 4\,K. At this temperature, the maximum magnetic moment of 1.97 ${\mu}_\mathrm{B}$ is obtained from the analysis which is close to the value (2.2 ${\mu}_\mathrm{B}$) estimated by Troyanchuka \emph{et al.}\cite{Troyanchuka}. Although ratio of magnetic to structural peak from the diffraction data of Troyanchuka \emph{et al.} indicate an overestimation of the magnetic moment.   In addition, the refinement also put an upper limit of  0.3 ${\mu}_\mathrm{B}$ on ferromagnetic component along the $x$ direction. Value of  0.3 ${\mu}_\mathrm{B}$ is one order of magnitude higher compare to moment arising due to D-M interaction between canted spins from Fe/Mn ions. 
Nd$^{3+}$ spins should order as ($F_x, C_y$) in the ${\Gamma}_{2}$ representation of $M$ atoms.   
However, the contribution of Nd$^{3+}$ sublattice moments could not be resolved from powder neutron diffraction pattern.\\
  \begin{figure}[h!] \center
       \begin{picture}(250,180)
        \put(0,-5){\includegraphics[width=240\unitlength]{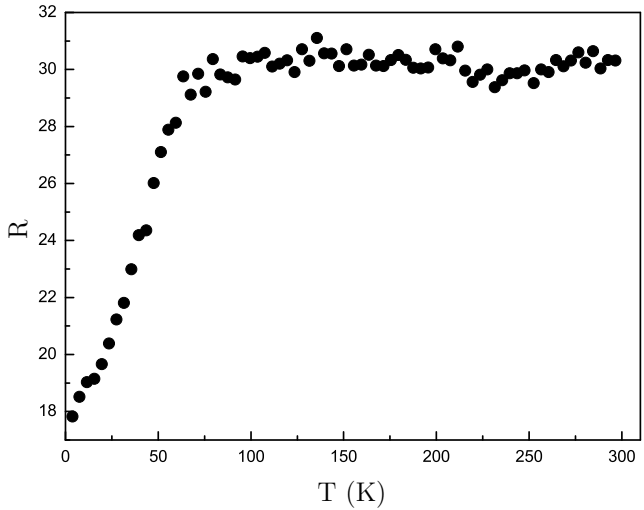}}
          
                    
       \end{picture}
        \caption{Temperature variation of flipping ratio (R) under an applied field of 50\,Oe for NFMO.}
        \label{ND_depol}
      \end{figure}
As discussed earlier, neutron diffraction and magnetization measurements show the presence of ferromagnetic component below 70\,K. To establish the presence of ferromagnetic component further, depolarization of incident polarized neutrons transmitted through the sample was checked. It may be noted that the incident neutron beam should not depolarize for antiferromagnetic ordering.  Fig.~\ref{ND_depol} shows temperature dependence of the flipping ratio. It remains constant down to $\sim$75\,K , below which it decreases slowly.  The observed decrease in the flipping ratio confirms the presence of ferromagnetic correlations, as observed in the magnetization and neutron diffraction study.  The slow decrease in the flipping ratio with temperature agrees well with neutron diffraction measurements where the volume fraction of the magnetic phase ${\Gamma}_{2}$ increases with decreasing temperature. It means volume fraction of the ${\Gamma}_{2}$ phase having ferromagnetic component increases. \\
However, the large remanent magnetization and coercivities from M-H loops hint possible role of the ferromagnetic polarization of the Nd$^{3+}$ sublattice in addition to the M$^{3+}$ spins. An important anomaly we observe in our studies is that the absolute values of magnetic moments are much lower than the expected range of 3-4 ${\mu}_\mathrm{B}$. A possible reason for this could be due to presence of inherent disorder in the system arising from random distribution of Mn and Fe, which prevents a complete coherent ordering of spins thereby reducing the absolute values of moments. \\
From the neutron diffraction and depolarization measurements on NFMO discussed above we can conclude that there exists an antiferromagnetic G-type ordering below 250\,K in the ${\Gamma}_{1}$ irreducible representation with spins aligned along y-axis. Magnetic ordering in ${\Gamma}_{1}$ representation does not have any associated weak ferromagnetism due to canting of the spins. Spins start to reorient below 70\,K and a new magnetic phase with ${\Gamma}_{2}$ representation appears in combination with ${\Gamma}_{1}$. An associated weak ferromagnetism also appears in  \textcolor{green}{${\Gamma}_{2}$} phase due to canting of spins. This is very clearly evident from the magnetization and depolarization measurements. Spin reorientation transition completes below 20\,K and only one phase with ${\Gamma}_{2}$ ($F_xG_z$) representation exists having $G_z$ antiferromagnetic ordering and a weak ferromagnetic component.\\
\section{Electronic structure}
\label{Elect_struct}
Due to complex magnetic behavior arising out of the interplay of Mn$^{3+}$ and Fe$^{3+}$ ions as evident from the experimental observations discussed above, we attempt to evaluate the ground state magnetic order of the system and the associated spin reorientation phenomenon in a systematic manner using first principles density functional theory calculations. Complete random distribution of Mn and Fe ions within the orthorhombic unit cell of NFMO will require a very large supercell which will be computationally very expensive. Therefore, we have considered three possible superlattices of Mn and Fe ions which are labelled as  (001), (110) and (111)-cationic ordering. Fig.~\ref{unit_cells} depicts these three structures. In the (001) ordering, the Mn and Fe planes are alternately arranged along the $c$ direction. In the (110) arrangement, Mn and Fe ions are alternately arranged in the $ab$-plane and the same arrangement is repeated along the c-direction, while in the (111) arrangement Mn and Fe ions are alternately arranged along all the three directions. 
  \begin{figure}[h!] \center
       \begin{picture}(250,90)
        \put(0,-5){\includegraphics[width=240\unitlength]{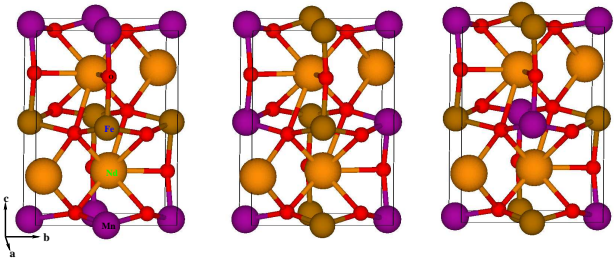}}
       \end{picture}
        \caption{Three possible superlattices of Fe and Mn ions considered in our calculations: (a) 001, (b) 110 and (c) 111 respectively.}
        \label{unit_cells}
      \end{figure}
For each of these three cationic arrangements, structural optimisation of the orthorhombic unit cell have been performed within GGA where the ionic positions were relaxed keeping lattice parameters fixed at their experimental values. These three structures were further optimised for various collinear magnetic orderings viz ferromagnetic (FM), A, C and G-type antiferromagnetic orderings within the unit cell. 
Since our experimental studies show a G-type magnetic ordering, the structural parameters have been listed in Table~\ref{table2} obtained for the three cationic orderings only for the G-type magnetic ordering. Experimentally, Mn-O and Fe-O bond lengths are found to be equal whereas in our relaxed structure, we obtain different bond-lengths for Mn-O and Fe-O. The MnO$_{6}$ shows maximum inequality in the bond lengths in the case of (001) cationic ordering similar to NdMnO$_{3}$. Thus among the three possible Mn/Fe arrangements, (001) shows maximal effect of J-T distortion at Mn site.
\begin{table}
\caption{Relaxed structural parameters of NFMO with G-type magnetic ordering within GGA. }
\begin{center}
\begin{tabular}{p{4.0cm}c c c c c c }\hline\hline 
Mn/Fe arrangement & (001) &  (111) &(110)\\ \hline 
Fe-O1 & 2.0423& 2.0349  & 1.9826  \\
Fe-O2 & 1.9964 & 2.0657 & 2.05558 \\ 
Fe-O2 & 2.0250  & 2.0358 & 2.0838  \\ 
Mn-O1 & 1.93196  & 1.9407 & 1.9274  \\ 
Mn-O2 & 1.92113 &1.9501  &1.9478 \\ 
Mn-O2 & 2.10256 & 1.9825 & 1.9592 \\   
\hline\hline
 
\end{tabular} 
\end{center}
\label{table2}
\end{table}

With the optimized structural parameters, GGA+U calculations for various U values are performed. The relative energies of three cationic (Fe/Mn) arrangements along with four magnetic ordering within GGA+U (U=6.95 eV, J=0.95eV) are listed in Table~\ref{table2} considering FM order in (001) structure to be at zero. Overall energy comparision shows that the G-type magnetic order (for (001) Fe/Mn arrangement) is the ground state which is consistent with our experimental observations discussed in previous sections.  
\begin{table}
\caption{ Relative energies (in meV) for various magnetic configurations within GGA+U (U=6.95 eV and J=0.95 eV) for the three possible Mn/Fe cationic ordering. }
\begin{center}
\begin{tabular}{p{4.0cm}c c c c c c }\hline\hline 
Magnetic ordering & (001) &  (110) &(111)\\ \hline 
FM & 0 & 349  & 687  \\
A &  -71 & 450 & 229 \\  
C & -145 & 187 & -107  \\ 
G & -458 & 298 & 50 \\  
\hline\hline
 
\end{tabular}
\end{center}
\label{table2}
\end{table}	
We now discuss the role of Coulomb correlation $U$ on the ground state electronic structure and magnetism for all the three cationic ordering. 
Using the relaxed structural coordinates, the self-consistent calculations were performed for various values of $U_\mathrm{eff}=U-J$ in the range 0 to 6\,eV. The values of $U$ and $J$ were kept identical for Mn and Fe. In absence of correlation, the Mn and Fe magnetic moments are observed to be much smaller than 4\,${\mu}_\mathrm{B}$ and 5\,${\mu}_\mathrm{B}$ expected from Mn$^{3+}$ and Fe$^{3+}$ respectively for all the magnetic orderings. For instance, in the G-type magnetic ordering, Mn and Fe show magnetic moment values of 3.2\,${\mu}_\mathrm{B}$ and 3.6\,${\mu}_\mathrm{B}$ respectively. This can be due to the hybridization between the $3d$ states of Mn and Fe with O $2p$ states. Furthermore, in the absence of correlations, the density of states also show a half metallic behavior for all the cationic and magnetic orderings. 
On incorporating Coulomb correlation, the ground state magnetic configuration undergoes a variation which also depends on the Mn/Fe ordering in the unit cell. In the case of (001)- arrangement, the G-type magnetic ordering emerges as the minimum energy state till a maximum value of $U_\mathrm{eff}$ = 6\,eV. In the case of (111) arrangement, it is found that till $U_\mathrm{eff}$ = 2.5\,eV, the G-type magnetic ordering remains as ground state. However with further increase in $U_\mathrm{eff}$, the C-type magnetic ordering becomes the minimum energy state. In the case of (110) cationic arrangement, the C-type always remains the ground state for all values of $U_\mathrm{eff}$ with the G-type remaining in close proximity. Thus the role of correlations in stabilizing the G-type magnetic ordering occurs probably for smaller values of $U_\mathrm{eff}$. This is unlike the theoretically studied double perovskite Ho$_{2}$MnFeO$_{6}$, in which though the correlations affect structural and electronic behaviour, it does not affect the magnetic ordering.\\ 
Fig.~\ref{DOS} shows the partial spin polarized density of states(DOS) of Mn and Fe for the three Fe/Mn arrangements with $U_\mathrm{eff}$=6\,eV. 
  \begin{figure}[b!] \center
       \begin{picture}(250,270)
        \put(0,-5){\includegraphics[width=235\unitlength]{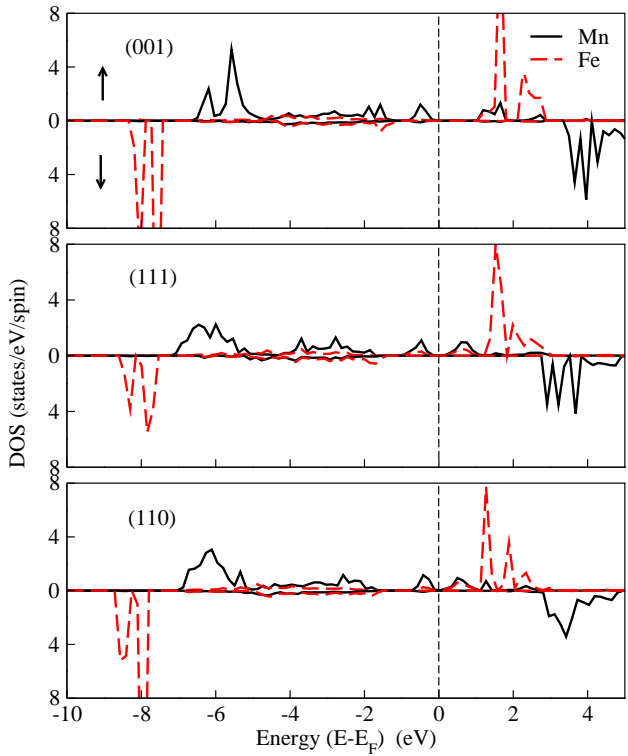}}
       \end{picture}
        \caption{Spin polarized DOS of Mn and Fe for the three cationic orderings within GGA+U (U=6.95\,eV and J=0.95\,eV).}
        \label{DOS}
      \end{figure}
It is clear from the DOS  that Mn ions exist in high spin 3+ valence state. $e_g$ states of Mn  split around Fermi level due to J-T effect as discussed above. On the other hand Fe ions also remain in high spin 3+ valence state with a large exchange splitting between majority and minority states pushing the fully filled majority spin states to an energy range far below (${\sim}$ 8\,eV) Fermi energy. Therefore, overlap between Mn and Fe states is seen to be very less and the band gap is solely decided by the J-T splitting of Mn $e_g$ states. The band gap is observed to be highest (\emph{i.e.} around 1\,eV)  for (001) ordering whereas it is around 0.3\,eV for (111). The (110) ordering gives rise to a very small gap. In case of (001) ordering, a co-operative J-T effect exists in addition to the local J-T effect in the Mn planes which gives rise to a long range orbital ordering in Mn plane similar to NdMnO$_3$ (i.e. alternate arrangement of d$_\mathrm{3x^2-r^2}$ and d$_\mathrm{3y^2-r^2}$ at neighboring Mn sites). This orbital ordering gives rise to an enhanced band gap in this case. It is also interesting to note at this point that due to very less overlap between Fe and Mn states in all the three ordering, the exchange interaction between Fe and Mn would be weaker in comparison to Mn-Mn and Fe-Fe exchange in NFMO. This is one possible reason for (001) (where Mn and Fe ions are segregated in separate $ab$-planes along $c$-direction) to come out as the lowest energy state.\\
Finally in order to the understand the mechanism of experimentally observed spin reorientation from high temperature $\Gamma_1$ phase to low temperature $\Gamma_2$ phase, we carried out calculations with non-collinear arrangement of spins and considering the spin-orbit interaction within GGA+U+SO approximation. For the high temperature $\Gamma_1$ phase, the corresponding experimental structure with optimized ionic positions was considered. Nd $4f$ moments were kept frozen in the core states and Fe/Mn spins considered to form a G-type magnetic ordering among themselves. The total energies for the three situations were calculated with Fe/Mn spins pointing along $a$, $b$ or $c$ directions.
For both (111) and (001) arrangements of Fe/Mn, $b$ direction was obtained as preferred (or easy) axis of magnetization. This result explains our experimental observation of $\Gamma_1$ magnetic phase below $T_\mathrm{N}$ and above 75\,K (high temperature phase). Moreover, this easy axis of magnetization along $b$ is further stabilized energetically by incorporation of Coulomb correlation $U$. In Fig.~\ref{non_collin}(a), the total energies for different directions have been shown for  (111) cationic ordering. In order to understand the role of Mn, the corresponding calculations for NdFeO$_3$ have been performed as well. In this case, it has been observed that the easy changes to $a$ as verified experimentally earlier in case of NdFeO$_3$ high temperature phase (previously reported by Chen et. al.\cite{Chenjap12}). 
  \begin{figure}[h!]
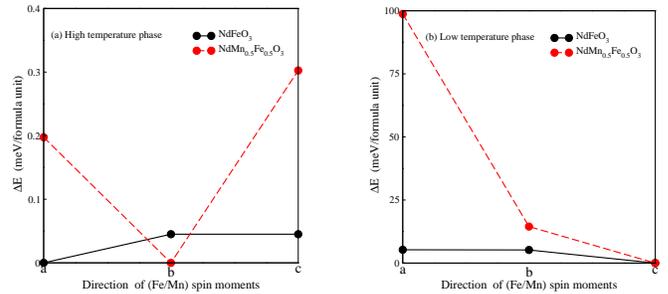
 \center
       \begin{picture}(250,110)
        \put(0,-5){\includegraphics[width=110\unitlength]{HT.eps}}
             \put(135,-5){\includegraphics[width=110\unitlength]{LT.eps}}
       \end{picture}
        \caption{Comparision of total energies for Fe/Mn spins pointing in $a$, $b$ and $c$ directions in case of NFMO and NdFeO$_3$ for (a) high temperature phase and (b) low temperature phase respectively.}
        \label{non_collin}
      \end{figure}
Similarly, for the low temperature $\Gamma_2$ phase (below spin reorientation transition) with G$_z$-type ordering, we have considered the corresponding experimental structure with optimized ionic positions and and Fe/Mn spins forming a G-type magnetic ordering among themselves. However, in this case the $4f$ electrons of Nd have been treated as valence electrons and their moments taken along $b$ direction as observed experimentally in case of NdFeO$_3$. Our results for NFMO and NdFeO$_3$ for low temperature phase have been presented in Fig.~\ref{non_collin} (b). It is clear from the calculations that $c$ direction is preferred direction for this low temperature  phase which is again consistent with our experimental observation of  $\Gamma_2$ (G$_z$, F$_x$) phase. This was also found to be the case for NdFeO$_3$\cite{Chenjap12}.
Above mentioned results indicate that Nd-Fe or Nd-Mn exchange interactions play dominant role in deciding the spin orientations of Fe/Mn spins in low temperature $\Gamma_2$ phase. In the high temperature $\Gamma_1$ phase, however, the Nd moments are not ordered (net Nd moments is zero) and hence Nd-Fe/Mn exchange is absent. In this situation, magnetic anisotropy is dictated by the orbital anisotropy of Mn $d$-orbitals. 

\section{Conclusions and Summary}
\label{Summ}
In summary, NFMO polycrystalline samples have been prepared and investigated in detail in order to understand its complex structural, magnetic and electronic properties.
The samples crystallize in the space group P$bnm$ with both Fe and Mn, occupying crystallographically equivalent positions and thus being randomly distributed in the crystal. Presence of local Jahn-Teller distortion at substituted Mn sites was evident from the change in measured lattice parameters and Mn(Fe)-O bond lengths. In addition, values of various other distortion parameters are also observed to be higher in comparison to their corresponding values for orthoferrite NdFeO$_3$, though these values are an order of magnitude smaller than that of NdMnO$_3$.\\
Magnetic properties of NFMO are similar to NdFeO$_3$ although with many interesting differences. The antiferromagnetic transition occurs at relatively low temperature (at 250\,K) in comparison to NdFeO$_3$, indicating a reduction in the strength of the various exchange interactions due to Mn. Our neutron diffraction data indicate presence of short range magnetic correlations even at 300\,K, while the full ordering develops at 250\,K. Unlike NdFeO$_3$, irreducible representation of magnetic symmetry is $\Gamma_1$ below $T_\mathrm{N}$ with spins aligned along $b$ direction in G-type antiferromagnetic structure. This is highly unusual, since this magnetic representation develops at lower temperatures below spin reorientation transition in most of the orthoferrites.Ê
The origin of this representation is attributed to the large single ion anisotropy of the Mn$^{3+}$ ions which is an order of magnitude larger than that of the Fe$^{3+}$ ion. However, a competing interaction start to become prominent between the rare earth spins and transition metal spins which results in a complete reorientation of the Mn(Fe) spins. The reorientation occurs between 75\,K and 25\,K with a coexistence of the two irreducible representations, $\Gamma_1$ and $\Gamma_2$. While the volume fraction of $\Gamma_2$ increases, there occurs a corresponding decrease in volume fraction of  $\Gamma_1$. At the lowest temperature, the magnetic structure entirely belongs to the $\Gamma_2$ representation with G-type antiferromagnetic structure and spin aligned along $c$ axis. In addition the ferromagnetic component of the spin moment ($F_x$) develops in $\Gamma_2$ phase with a upper limit of 0.3\,$\mu_\mathrm{B}$ obtained from our Rietveld analysis. This is also concomitant with the coercivity obtained from the M-H curves which attain a minimum value in the reorientation region. We also found evidence of ferromagnetic correlations from our neutron depolarization measurements. However, a ferromagnetic moment of 0.3\,$\mu_\mathrm{B}$ is a rather large value than that is expected solely from the antisymmetric interactions between Mn and Fe. This indicates a definite role of Nd$^{3+}$ spins in polarizing the Mn and Fe spins at lower temperatures. However, enough experimental evidence on the ordering of Nd moments is lacking at present probably due to short and distributed length scale.\\
 The process of reorientation is explained by our first principles density functional theory calculations considering non collinear spin arrangements within GGA+U+SO approximation. From our non-collinear calculations we observe that in the absence of any net Nd moment, the preferred direction of Fe/Mn spins is $b$ with G-type arrangement for high temperature phase (below $T_\mathrm{N}$). And the alignment of spins changes to the $c$-axis for low temperature phase (below spin reorientation transition) after inclusion of Nd spins. This suggests that its the Nd $4f$-Mn(Fe) $3d$ interaction which plays a role in reorientation of the spins at low temperature. Thus the results from our first principles calculations 
corroborates further our experimental observations in understanding the complex 
spin reorientation process in this compound.
\section{Acknowledgement}
This work was supported by the UGC-DAE Consortium for Scientific Research through grant number CRS-M-228. AS and AR acknowledge MHRD and CSIR (India) respectively for research fellowships.
\bibliography{NFMO_Neutron}
\end{document}